\documentclass[conference, letter]{IEEEtran}

\usepackage{amsmath}
\usepackage{amssymb}
\usepackage{latexsym}
\usepackage{verbatim}
\usepackage{subfigure}
\usepackage[final]{graphicx}
\usepackage{psfrag}
\usepackage{hyperref}

\usepackage{tcolorbox}
\definecolor{darkred}{RGB}{250,0,0}
\definecolor{darkgreen}{RGB}{0,150,0}
\definecolor{myblue}{RGB}{0,0,250}
\definecolor{darkblue}{RGB}{0,0,200}
\hypersetup{colorlinks=true, linkcolor=darkred, citecolor=myblue, urlcolor=darkblue}

\newtheorem{theorem}{Theorem}

{\begin{list}               
    {$\bullet$ \hfill}{
        \setlength{\leftmargin}{\parindent}
        \setlength{\parsep}{0.04\baselineskip}
        \setlength{\itemsep}{0.5\parsep}
        \setlength{\labelwidth}{\leftmargin}
        \setlength{\labelsep}{0em}}
    }
{\end{list}}

\providecommand{\eref}[1]{{\eqref{eq:#1}}}  
\providecommand{\cref}[1]{Chapter~\ref{chap:#1}}
\providecommand{\sref}[1]{Section~\ref{sec:#1}}
\providecommand{\fref}[1]{Figure~\ref{fig:#1}}

\providecommand{\R}{\ensuremath{\mathbb{R}}}

\providecommand{\abs}[1]{\lvert#1\rvert}
\providecommand{\norm}[1]{\lVert#1\rVert}

\providecommand{\set}[1]{\left\{#1\right\}}

\providecommand{\bydef}{\overset{\text{def}}{=}}

\renewcommand{\vec}[1]{\ensuremath{\boldsymbol{#1}}}
\providecommand{\mat}[1]{\ensuremath{\boldsymbol{#1}}}

\providecommand{\mA}{\mat{A}} 
\providecommand{\mC}{\mat{C}}

\providecommand{\mI}{\mat{I}}

 \providecommand{\mG}{\mat{G}}

\providecommand{\va}{\vec{a}} \providecommand{\vb}{\vec{b}}
\providecommand{\vc}{\vec{c}} \providecommand{\ve}{\vec{e}}
\providecommand{\vh}{\vec{h}}

\providecommand{\vq}{\vec{q}} 
 
\providecommand{\vg}{\vec{g}}
\providecommand{\vu}{\vec{u}} \providecommand{\vw}{\vec{w}}
\providecommand{\vx}{\vec{x}} 
\providecommand{\vz}{\vec{z}} 
 
 \providecommand{\vv}{\vec{v}}

\providecommand{\vmu}{\vec{\mu}}
\providecommand{\vlambda}{\vec{\lambda}}
\providecommand{\veta}{\vec{\eta}}

\newcommand{\vsp}{\vspace{3pt}}

\usepackage{lipsum}    
\usepackage{ifthen}
\usepackage{tcolorbox}
\newboolean{showcomments}
\setboolean{showcomments}{true}
\newcommand{\oussama}[1]{\ifthenelse{\boolean{showcomments}}
{ \textcolor{red}{(Oussama says:  #1)}}{}}
\newcommand{\christos}[1]{\ifthenelse{\boolean{showcomments}}
{ \textcolor{blue}{(Christos says: #1)} } {} }
\newcommand{\yue}[1]{\ifthenelse{\boolean{showcomments}}
{ \textcolor{magenta}{(Yue says:  #1)}}{}}
\usepackage{cite}

\usepackage{color}
\usepackage{amsmath}

\usepackage{booktabs,multirow}

\usepackage{enumitem}
\usepackage{commath}

\usepackage{algorithm, algorithmic}

\usepackage{esdiff}

\usepackage{pgf,tikz,psfrag}

\usepackage{dsfont}

\DeclareMathOperator{\atan}{atan}

\providecommand{\vxi}{\vec{\xi}}

\begin{document}
  
\title{Phase Retrieval via Linear Programming: Fundamental Limits and Algorithmic Improvements}

\author{\IEEEauthorblockN{Oussama Dhifallah$^1$, Christos Thrampoulidis$^2$ and Yue M. Lu$^1$}
\IEEEauthorblockA{$^1$Harvard University, Cambridge, MA 02138, USA\\
{Email: oussama\_dhifallah@g.harvard.edu, yuelu@seas.harvard.edu}
}
\IEEEauthorblockA{$^2$Massachusetts Institute of Technology, Cambridge, MA 02139, USA\\
{Email: cthrampo@mit.edu}}
}

\maketitle

\begin{abstract}
A recently proposed convex formulation of the phase retrieval problem estimates the unknown signal by solving a simple linear program. This new scheme, known as PhaseMax, is computationally efficient compared to standard convex relaxation methods based on lifting techniques. In this paper, we present an exact performance analysis of PhaseMax under Gaussian measurements in the large system limit. In contrast to previously known performance bounds in the literature, our results are asymptotically exact and they also reveal a sharp phase transition phenomenon. Furthermore, the geometrical insights gained from our analysis led us to a novel nonconvex formulation of the phase retrieval problem and an accompanying iterative algorithm based on successive linearization and maximization over a polytope. This new algorithm, which we call PhaseLamp, has provably superior recovery performance over the original PhaseMax method.

\end{abstract}


\section{Introduction}\label{sec:intro}
\subsection{Background}\label{sec:back}

We consider the real-valued phase retrieval problem, which seeks to recover an unknown signal vector $\vxi\in\R^n$ from  $m$ magnitude measurements $\lbrace y_i, 1\leq i \leq m \rbrace$ of the form:
\begin{equation}\label{eq:abs}
y_i = \abs{\va_i^T \vxi},
\end{equation}
where  $\lbrace \va_i\in\R^n, 1 \leq i \leq m \rbrace$ is a set of (known) sensing vectors. In order to recover $\vxi$, we must resolve the uncertainty due to the missing phase (or sign) information. This is a classical problem \cite{Gerchberg:1972jk, Fienup:82}, with many applications in applied physics and engineering. Over the past decade, it has attracted significant attention in the optimization and signal processing communities, with a particular effort towards establishing rigorous recovery guarantees for either already existing or newly proposed solution methods. See \cite{Candes:2013xy, Jaganathan:2013zl, Waldspurger:2015rz, Netrapalli:2013qv, Candes:2015fv, WangGY:2016} and many references therein.

Among the most well-established methods are those based on semidefinite relaxation (\emph{e.g.}, \cite{balan2009painless,Candes:2013xy}.) Such convex optimization methods operate by lifting the original $n$-dimensional natural parameter space to a higher dimensional matrix space. Unfortunately, the increase in the dimensionality introduces challenges in computational complexity and memory requirement for the resulting algorithms. Subsequent works \cite{Netrapalli:2013qv,Candes:2015fv, Chen:2015eu, WangGY:2016} suggest going around this issue by developing nonconvex formulations of the phase retrieval problem and solution algorithms that start with a careful spectral initialization \cite{Netrapalli:2013qv, Chen:2015eu, LuL:17}, which is then iteratively refined by a gradient-descent-like scheme of low computational complexity. 

More recently, an alternative convex formulation of the phase retrieval problem in the original $n$-dimensional parameter space was independently proposed by two groups of authors \cite{phmax2, phmax}. The resulting method, referred to as PhaseMax in \cite{phmax}, relaxes the \emph{nonconvex equality} constraints in \eref{abs} to \emph{convex inequality} constraints, and solves the following linear program:
\begin{equation}\label{eq:lp_form}
\begin{aligned}
\widehat{\vx}&=\underset{{\vx}\in\mathbb{R}^n}{\arg\,\max}~~~ {\vx}_\text{init}^{T}\,{\vx}\\	
&~~~~~~~~\text{s.t.}~~~~  \abs{\va_i^T \vx} \leq y_i, \text{ for }  1 \le i \le m.
\end{aligned}
\end{equation} 
Here, $\vx_\text{init}$ represents an initial guess (or ``anchor vector'') that is correlated with the target vector $\vxi$.

Despite its simple formulation, PhaseMax has strong theoretical performance guarantees. Existing analysis \cite{phmax2, phmax, Hand:2016cs} shows that PhaseMax achieves exact signal recovery from a nearly optimal number of \emph{random measurements}. Specifically, in the case when the sensing vectors are drawn from the Gaussian distribution, the required number of measurements for perfect reconstruction is shown to be linear with respect to the underlying dimension, \emph{i.e.}, $m = c\,n$ for some constant $c$ that depends on the quality of the initial vector $\vx_\text{init}$. The analysis in \cite{phmax2, phmax, Hand:2016cs} gives various upper bounds on the constant $c$. The exact value of $c$, namely the sharp \emph{phase transition threshold}, is predicted in a recent work \cite{Ouss:17} by a subset of the authors of the current paper, but the analysis in \cite{Ouss:17} uses the non-rigorous replica method from statistical physics.

%

\subsection{Contributions}

Our main contributions in this paper are two-fold.

\noindent\emph{1. Exact recovery guarantees.} We present an exact performance analysis of the PhaseMax method for the (real-valued) phase retrieval problem with Gaussian sensing vectors in the large system limit. When  $m,n\rightarrow\infty$ at a proportional ratio $\alpha =m/n$, we \emph{rigorously} establish the exact phase transition threshold.
Furthermore, in the regime where perfect recovery is not feasible, we derive asymptotically exact formulas for the normalized mean squared error (NMSE), defined as 
\[
\text{NMSE}_n \bydef {{\min\{\norm{\vxi - \widehat\vx}_2^2, \norm{\vxi + \widehat\vx}_2^2\}}}/\,{{\norm{\vxi}_2^2}}.
\]
Our formulas reveal the precise dependence of the NMSE on the oversampling ratio $\alpha$ and on the quality of the initial guess $\vx_\text{init}$ as measured via the input cosine similarity 
\begin{equation}\label{eq:cosin}
\rho_{\text{init}} \bydef \frac{\abs{\vx_\text{init}^T \vxi}}{\norm{\vx_\text{init}}_2 \norm{\vxi}_2}.
\end{equation}
Our main results can be summarized by the following asymptotic characterization of the NMSE:
\begin{equation}\label{eq:pt_in}
\mathrm{NMSE}_n \xrightarrow[]{n\to\infty} \begin{cases}
0  &\hspace{-2.5mm},\text{if}~\rho_\text{init} > \rho_c(\alpha),\\
f(\rho_{\text{init}},\alpha)>0 &\hspace{-2.5mm},\text{otherwise},
\end{cases} 
\end{equation} 
where 
\begin{equation}\label{eq:ptt}
\rho_c(\alpha)\bydef\sqrt{1- \frac{{\pi}/{\alpha}}{\tan({\pi}/{\alpha} )}},
\end{equation}
 and $f(\rho_{\text{init}},\alpha)$ is explicitly determined by solving a one-dimensional deterministic fixed point equation [see \eref{fconprob} and Theorem~\ref{thm:the1}.]

Our analysis builds upon the recently developed convex Gaussian min-max theorem (CGMT) \cite{chris:151, chris:152}, which involves a tight version of a classical Gaussian comparison inequality\cite{Gordon:85}. The CGMT framework has been successfully applied to derive precise performance guarantees for structured signal recovery under (noisy) linear Gaussian measurements, \emph{e.g.}, \cite{Sto:13, thrampoulidis2015asymptotically, chris:151, chris:152}. In \cite{thrampoulidis2015lasso}, the CGMT is used to study signal recovery from a class of non-linear measurements.  However, this excludes magnitude-only or quadratic measurements that are relevant for the phase retrieval problem considered here. 

The precise nature of our results serves to tighten up the previously known performance bounds of PhaseMax \cite{phmax2, phmax, hand2016elementary}. They also exactly match and thus rigorously verify the predictions in \cite{Ouss:17} obtained from the non-rigorous replica method from statistical physics. In fact, to the best of our knowledge, this is the first \emph{exact} performance analysis of any of the existing solution methods for the phase retrieval problem.
 
\vspace{1ex}

\noindent\emph{2. From precise analysis to algorithmic improvements.} As the second contribution of this paper, we propose a new nonconvex formulation and an efficient iterative algorithm for the phase retrieval problem. Our new formulation is inspired by PhaseMax, the key idea of which is to relax the nonconvex equality constraints in \eref{abs} to convex \emph{inequality} constraints. The intersections of all these inequality constraints form a high-dimensional (random) polytope. Our analysis of the PhaseMax method provides useful insights on the exact high-dimensional geometry of that random polytope. These insights then lead us to a novel nonconvex formulation of the phase retrieval problem, as follows: 
\begin{equation}\label{eq:qb_form}
\begin{aligned}
\widehat{\vx}&=\underset{{\vx}\in\mathbb{R}^n}{\arg\,\max}~~~ \norm{\vx}^2_2\\	
&~~~~~~~~\text{s.t.}~~~~  \abs{\va_i^T \vx} \leq y_i, \text{ for }  1 \le i \le m.
\end{aligned}
\end{equation}
Note that \eref{qb_form} is indeed a nonconvex problem, as we aim to \emph{maximize} a convex function over a convex domain. We devise an efficient iterative method, which we call \emph{PhaseLamp}, to solve \eqref{eq:qb_form}. The name comes from the fact that the algorithm is based on the idea of successive linearization and maximization over a polytope, where in each step we solve a PhaseMax problem with the initialization given by the estimate from the previous iteration.

\begin{figure}[t]
    \centering
    \includegraphics[width=0.3\textwidth]{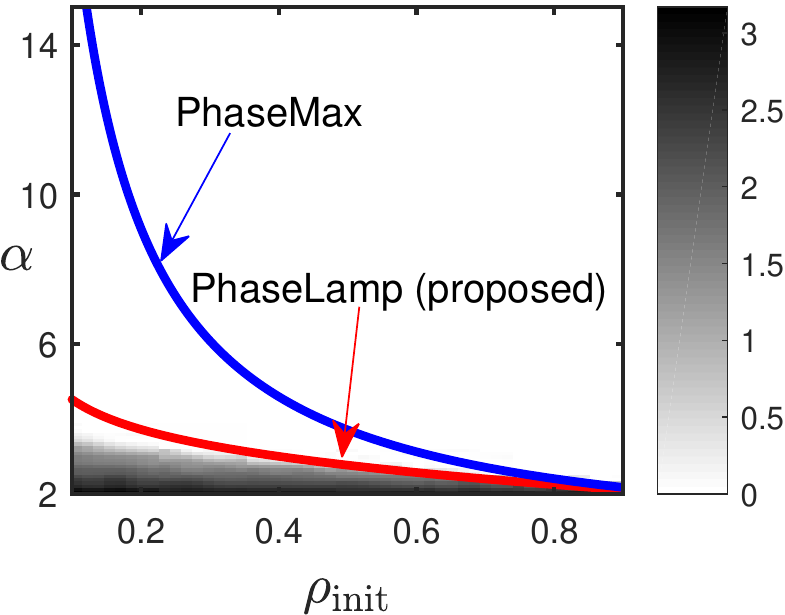}
    \caption{The NMSE of the PhaseLamp method: theory versus simulations. The signal dimension is set to $n=1000$, and the results are averaged over $10$ independent trials. The red curve shows the sufficient condition, as given in \eref{slam_meth_in}, for PhaseLamp to successfully recover the target signal. This is compared against the blue curve, which shows the phase transition boundary of the original PhaseMax method as given in \eref{ptt}. }
    \label{fig:phase}
\end{figure}

We prove that the proposed PhaseLamp method has (strictly) superior recovery performance over PhaseMax. Specifically, we show that a \emph{sufficient condition} for PhaseLamp to perfectly recover the target signal $\vxi$ is
\begin{equation}\label{eq:slam_meth_in}
\rho_\text{init} > \rho_s(\alpha),
\end{equation}
 where $\rho_s(\alpha)$ is determined explicitly by solving a one-dimensional deterministic fixed point equation (see \eref{slam_fp} and Theorem~\ref{the2}.) In particular, $\rho_s(\alpha)$ is strictly smaller than $\rho_c(\alpha)$ as defined in \eqref{eq:ptt}. This is illustrated through a numerical example shown in Figure \ref{fig:phase}. We can see that the proposed PhaseLamp method significantly improves the recovery performance of the PhaseMax method, especially in the more challenging, and arguably the more practically relevant regime of small input cosine similarities $\rho_\text{init}$. Moreover, although \eref{slam_meth_in} is only a sufficient condition, it nevertheless provides a good estimate of the actual performance of the algorithm. 
 



\section{Precise Performance Analysis of PhaseMax}
\label{sec:main_results}

\subsection{Technical Assumptions}
\label{sec:assp}
The asymptotic predictions derived in this paper are based on the following assumptions.
\begin{enumerate}[label={(A.\arabic*)}]
\item The sensing vectors $\set{\va_i}_{1 \le i \le m}$ are known and drawn independently from a Gaussian distribution with zero mean and covariance matrix $\mI_n$.

\item $m = m(n)$ with $\alpha_n = m(n) / n \rightarrow \alpha > 0$ as $n \rightarrow \infty$.

\item Both the target signal vector $\vxi$ and the initial guess vector $\vx_\text{init}$ are independent from the sensing vectors $\set{\va_i}_{1 \le i \le m}$.


\end{enumerate}

%

For convenience, we shall also assume that the initial guess vector $\vx_\text{init}$ has a \emph{positive} correlation with the target signal vector $\vxi$, \emph{i.e.}, $\vxi^T \vx_\text{init} > 0$. This can be made without loss of generality since the vectors $\vxi$ and $-\vxi$ are both valid target vectors.


\subsection{Fundamental Limits of PhaseMax}

In this section, we characterize the asymptotic NMSE of the PhaseMax method under the stated assumptions. In particular, our results point out necessary and sufficient conditions on the oversampling ratio $\alpha$ and on the {cosine similarity} $\rho_\text{init}$ for perfect recovery. 

In order to state our results we need a few definitions. For any fixed {cosine similarity} $\rho_\text{init}$ and fixed oversampling ratio $\alpha>2$, define $s^*$ as follows:
\begin{equation}\label{eq:fconprob}
\begin{aligned}
s^{\ast}&\bydef\underset{0 \leq {s} \leq 1}{\arg\,\max}\ \ {\rho_{\text{init}}} s+\sqrt{(1-\rho_{\text{init}}^2)g_{\alpha}(s)}
\end{aligned}
\end{equation} 
where the function $g_\alpha(s):[-1,1]\rightarrow(0,\infty)$ (parametrized by $\alpha$) is given by
\begin{equation}\label{eq:fun_g}
g_{\alpha}(s)\bydef-1-s^2+\frac{2\alpha \, r_{\alpha}(s)}{\pi}+\frac{2\alpha s}{\pi }\text{atan}\left( \frac{s}{r_{\alpha}(s)+c_{\alpha}} \right),
\end{equation}
with $c_{\alpha}=1/\text{tan}\left( \pi/\alpha \right)$ and 
\begin{equation}
r_{\alpha}(s)\bydef\sqrt{c_{\alpha}^2+1-s^2}-c_{\alpha}.
\end{equation}
Moreover, define
\begin{equation}\label{eq:rstar}
r^\ast\bydef r_{\alpha}(s^\ast).
\end{equation}

\vsp
\begin{theorem}[NMSE of PhaseMax]\label{thm:the1}
For any fixed input {cosine similarity} $\rho_\text{init} > 0$ and any fixed oversampling ratio $\alpha>2$, let $s^\ast, r^\ast$ be defined as in \eqref{eq:fconprob} and \eqref{eq:rstar}, respectively. Then, under the assumptions in \sref{assp}, 
the NMSE of the PhaseMax method converges in probability as follows:
\begin{align}\label{eq:anmse}
\mathrm{NMSE}_n \xrightarrow[]{n\to\infty} 1+(s^\ast)^2+(r^\ast)^2-2\abs{s^\ast}.
\end{align}
\end{theorem}

The proof of Theorem \ref{thm:the1} is based on the CGMT\cite{chris:151,chris:152}. To streamline our presentation, we postpone a sketch of the  proof to the appendix. Theorem \ref{thm:the1} accurately predicts the NMSE of PhaseMax in the large system limit. The prediction is expressed in terms of $s^\ast$, the solution to the one-dimensional deterministic maximization problem in \eqref{eq:fconprob}. It can be shown that this optimization problem is concave and that $s^\ast$ can be uniquely determined by a fixed point equation. 


\vsp
\begin{theorem}[Phase transition of PhaseMax]\label{thm:pt}
For any fixed {cosine similarity} $\rho_\text{init}$ and any fixed oversampling ratio $\alpha>2$, the PhaseMax method perfectly recovers the target signal (in the sense that $\mathrm{NMSE}_n \xrightarrow[]{n\to\infty} 0$, in probability) \emph{if and only if}
\begin{align}\label{eq:phb}
\rho_\text{init} > \sqrt{1- \frac{{\pi/\alpha}}{\tan({\pi}/{\alpha} )}} \bydef \rho_c(\alpha).
\end{align}

\end{theorem}
%
%
%

A sketch of the proof of Theorem~\ref{thm:pt} can be found at the end of the appendix. The theorem establishes a precise phase transition behavior on the performance of PhaseMax: for any fixed oversampling ratio $\alpha > 2$, there is a critical cosine similarity $\rho_c(\alpha)$ such that the PhaseMax method perfectly recovers the target signal vector $\vxi$ if and only if $\rho_\text{init} > \rho_c(\alpha)$. 

%

\subsection{Numerical Simulations}

In this section, we present simulation results that verify the validity of our predictions given in Theorems \ref{thm:the1} and \ref{thm:pt}. We solve the convex optimization problem \eref{lp_form} using the technique introduced in \cite{fasta}. The signal dimension is set to $n=1000$. 


\begin{figure}[h!]
    \centering
    \subfigure[]{\label{fig:nmse_1}
    \includegraphics[width=0.47\linewidth]{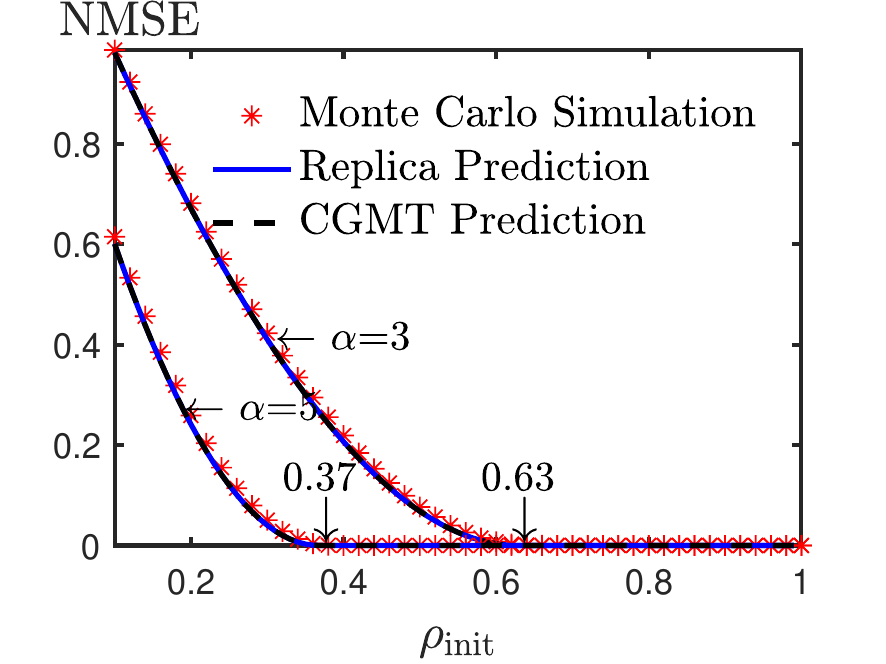}
    }
    \subfigure[]{\label{fig:nmse_2}
        \includegraphics[width=0.47\linewidth]{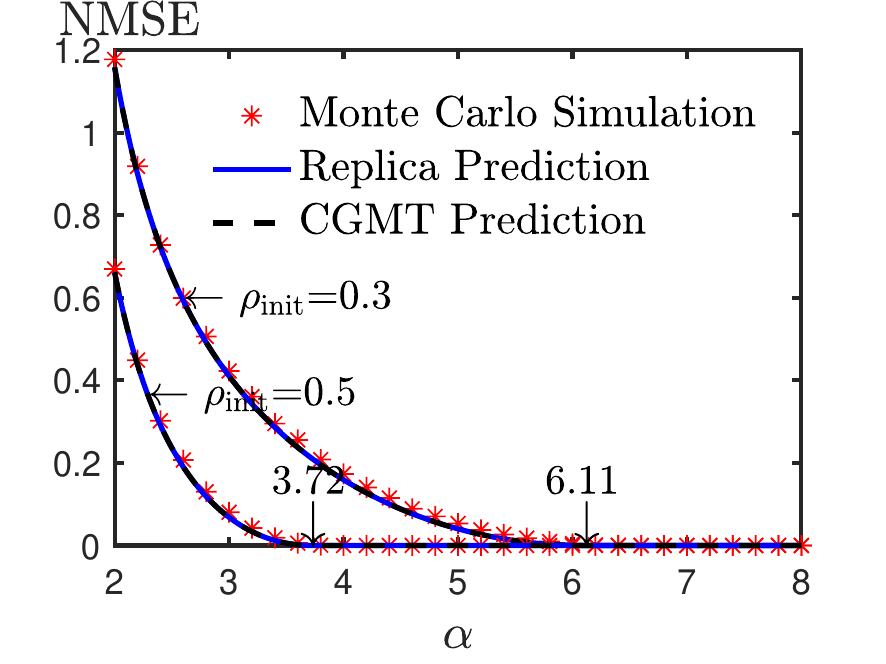}}

    \caption{Asymptotic predictions v.s. numerical simulations. (a) The NMSE of the PhaseMax method as a function of $\rho_\text{init}$, for two different values of $\alpha$; (b) The NMSE of the PhaseMax method as a function of $\alpha$, for two different values of $\rho_\text{init}$. The results are averaged over $50$ independent Monte Carlo trials. The CGMT expressions in this paper verify the non-rigorous replica predictions derived in \cite{Ouss:17}. The asymptotic formulas are also in excellent agreement with the actual performance the PhaseMax method, for $n = 1000$.}
        \label{fig1}
\end{figure}

In the first example, we investigate the performance of our asymptotic predictions for PhaseMax given in \eref{anmse}. Specifically, we compare the asymptotic predictions against simulation results for different values of the input cosine similarity $\rho_{\text{init}}$ and the oversampling ratio $\alpha$. \fref{nmse_1} illustrates the NMSE of PhaseMax as a function of the input cosine similarity given in \eref{cosin}, for two different values of the oversampling ratio. It can be noticed that our asymptotic predictions of the PhaseMax performance obtained using the CGMT perfectly match the asymptotic predictions derived using the non-rigorous replica method \cite{Ouss:17}. The asymptotic predictions are also in excellent agreement with the actual performance of the PhaseMax method in finite dimensions. \fref{nmse_1} further shows that the critical cosine similarity for the considered oversampling ratio $\alpha$ is given by $\rho_{\text{init}}(\alpha=3) \approx 0.63$ and $\rho_{\text{init}}(\alpha=5)\approx 0.37$, respectively, which perfectly matches the theoretical predictions given in Theorem~\ref{thm:pt}.

\fref{nmse_2} provides an additional example, where we plot the NMSE of the PhaseMax method as a function of the oversampling ratio, for two different values of the input cosine similarity. Again, the asymptotic performance obtained in our analysis perfectly matches the actual performance of the algorithm. 


\section{Algorithmic Improvements}
\label{sec:slam_alg}

\subsection{PhaseLamp: A New Algorithm}

This section proposes an efficient iterative algorithm to solve the norm maximization problem formulated in \eref{qb_form}. The optimization problem \eref{qb_form} consists of maximizing a convex function over a convex feasibility set. Hence, it is nonconcave where the cost function can be written as a difference of concave functions. To solve this problem, we propose the following scheme, named PhaseLamp, based on the idea of successive \emph{linearization and maximization over a polytope}:
\begin{equation}\label{eq:itphmax_form}
\begin{aligned}
{\vx}_{k+1}&=\underset{{\vx}}{\arg\,\max}~~~ \vx_k^T\vx\\	
&~~~~~~~~\text{s.t.}~~~~  \abs{\va_i^T \vx} \leq y_i, \text{ for }  1 \le i \le m.
\end{aligned}
\end{equation}
In essence, at each iteration we approximate (\emph{i.e.}, linearize) the cost function of \eref{qb_form} via 
\[
\norm{\vx}_2^2 \approx \vx_k^T \vx,
\]
where $\vx_k$ is the estimate obtained from the previous iteration. The proposed algorithm starts from an initial guess $\vx_0 =\vx_{\mathrm{init}}$ of the target vector $\vxi$. In practice, we terminate the proposed iterative algorithm when the number of iterations exceeds a pre-specified number $I_{\text{max}}$ or when $\norm{\vx_{k+1}-\vx_{k}}_2 \leq \epsilon$ for some fixed threshold $\epsilon > 0$.

There are several ways to interpret the proposed PhaseLamp algorithm. First, it can be viewed as an iterative and \emph{bootstrapped} version of the PhaseMax method \eref{lp_form} where at each iteration the previous optimal solution is used as an (improved) initial guess of the target signal vector $\vxi$. Second, PhaseLamp is a special case of a minorize-maximization (MM) algorithm \cite{Lange:16}. To see this, we note from the convexity of the cost function $\norm{\vx}_2^2$ that
\begin{equation}
\norm{\vx}_2^2 \geq \vx_k^T\vx_k+2\vx_k^T\left( \vx-\vx_k \right), \forall \vx_k,\vx \in\mathbb{R}^n. \label{eq:f_order_conv}
\end{equation}
The PhaseLamp procedure consists of iteratively maximizing the lower bound in \eref{f_order_conv} over the convex feasibility set in \eref{qb_form}. One particular property of the MM procedure is that it guarantees that the objective value of the optimization problem \eref{qb_form} is nondecreasing, \emph{i.e.},
\begin{equation}
\norm{\vx_{k+1}}_2^2 \geq \norm{\vx_k}_2^2, \forall k \geq 1.
\end{equation} 

Due to the nonconcavity of the maximization problem \eref{qb_form}, the proposed iterative algorithm is not guaranteed to converge to the global optimal solution of \eref{qb_form}. One particular property of the fixed points of the optimization problem \eref{itphmax_form} is that it is an extreme point of the feasibility set given in \eref{qb_form}.

\subsection{Performance Guarantees for PhaseLamp}
Using the analysis strategy that leads to Theorems \ref{thm:the1} and \ref{thm:pt}, we are further able to derive a sufficient condition for PhaseLamp to perfectly recover the target signal $\vxi$.

Again, we first need a few definitions. One can show that, for any $\alpha > 2$, the equation
\begin{equation}\label{eq:slam_fp}
\begin{aligned}
& \theta \cos^2\theta + (1 + 3 \sin^2\theta) \atan\left(\frac{\sin \theta \cos \theta}{1 + \sin^2 \theta}\right)\\
 &\qquad\qquad= 2 \sin \theta \cos \theta + \Big(\frac{\pi}{\alpha}\Big) \sin^2 \theta \cos^2 \theta,
\end{aligned}
\end{equation}
has a unique solution in the interval $\theta \in (0, \pi/2)$. We denote that solution by $\theta^\ast_\alpha$. Let
\begin{align}
\widehat{s}_\alpha\bydef \frac{\text{tan}(\theta^\ast_\alpha)}{\sqrt{1+c_{\alpha}^2+\text{tan}(\theta^\ast_\alpha)^{2}}+c_{\alpha}},
\end{align}
where $c_{\alpha}=1/\text{tan}\left( \pi/\alpha \right)$, and 
\begin{align}
\ell_\alpha \bydef\frac{\widehat{s}_\alpha-\frac{\alpha}{\pi}\text{atan}\left( \frac{\widehat{s}_\alpha}{\sqrt{ c_{\alpha}^2+1-\widehat{s}_\alpha^2} } \right)}{\sqrt{g_\alpha(\widehat{s}_\alpha)}},
\end{align}
where $g_\alpha$ is the function defined in \eref{fun_g}. We are now ready to state the main theorem of this section.

\begin{theorem}[Sufficient condition for perfect recovery]
\label{the2}
PhaseLamp perfectly recovers the unknown signal, \emph{i.e.}, it holds in probability that
$\mathrm{NMSE}_n \xrightarrow[]{n\to\infty}0$, if
\begin{align}\label{eq:slamp}
\rho_\text{init} > \frac{\ell_\alpha}{\sqrt{\ell^2_\alpha+1}} =: \rho_s(\alpha).
\end{align} 
\end{theorem}

The proof of Theorem~\ref{the2} is based on CGMT and the properties of the fixed points of the optimization problem \eref{itphmax_form}. Due to space constraint, we defer the proof of this theorem to the long version of the current paper. Unlike the asymptotically exact characterization given in Theorem~\ref{thm:pt}, the condition given in Theorem~\ref{the2} is sufficient but not necessary. However, as shown in \fref{phase} and the additional simulation results given in the next section, the condition in \eref{slamp} provides a reasonably tight bound on the actual performance of the PhaseLamp algorithm.

\subsection{Numerical Results}
\label{sec:numerical}

We present some numerical results to illustrate the performance of the proposed PhaseLamp method and our theoretical predictions given in Theorem~\ref{the2}. In our experiments, the signal dimension is set to $n = 1000$. \fref{fig2} plots the NMSE as a function of the oversampling ratio for two different values of the input cosine similarity ($\rho_\text{init} = 0.1$ and $\rho_\text{init} = 0.3$, respectively.) 

We observe that the proposed PhaseLamp method indeed outperforms the original PhaseMax method, and the amount of improvement is greater when the input cosine similarity is smaller. Specifically, for $\rho_\text{init} = 0.1$, the empirical minimum sampling ratio for PhaseLamp to perfectly recover $\vxi$ is at $\alpha \approx 3.3$, whereas PhaseMax requires $\alpha \approx 18.2$ . [The latter point is not shown in \fref{slam_phmax1}.] Moreover, \fref{fig2} also shows that the sufficient condition developed in Theorem~\ref{the2} for PhaseLamp provides a good estimate of the actual performance of the proposed algorithm. For example, \fref{slam_phmax2} demonstrates that the actual critical oversampling ratio of PhaseLamp is at $\alpha_c\approx2.9$, whereas the sufficient oversampling ratio as given in Theorem~\ref{the2} is $\alpha_s \approx 3.3$.



\begin{figure}[t!]
    \centering
    \subfigure[]{\label{fig:slam_phmax1}
    \includegraphics[width=0.47\linewidth]{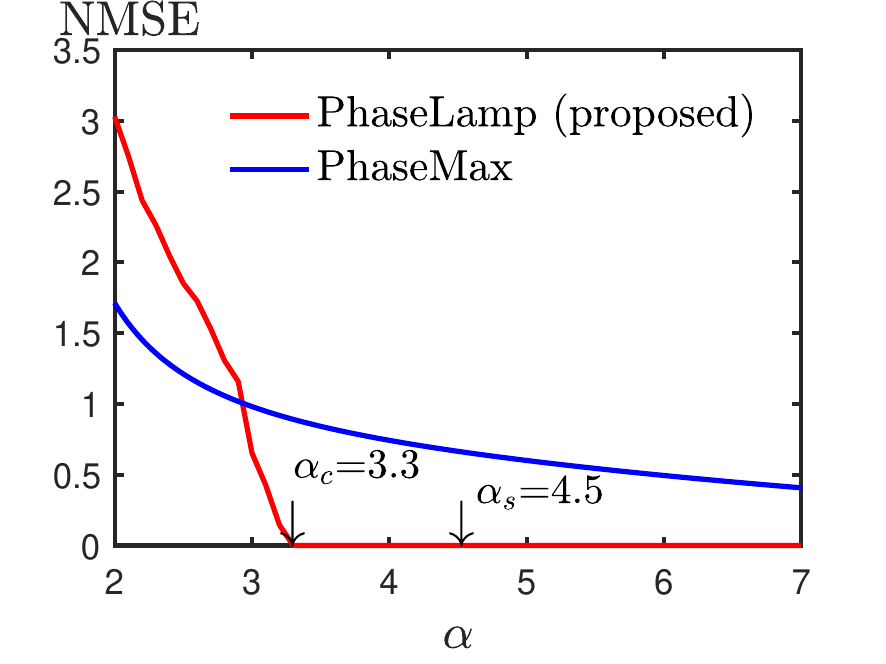}
    }
    \subfigure[]{\label{fig:slam_phmax2}
        \includegraphics[width=0.47\linewidth]{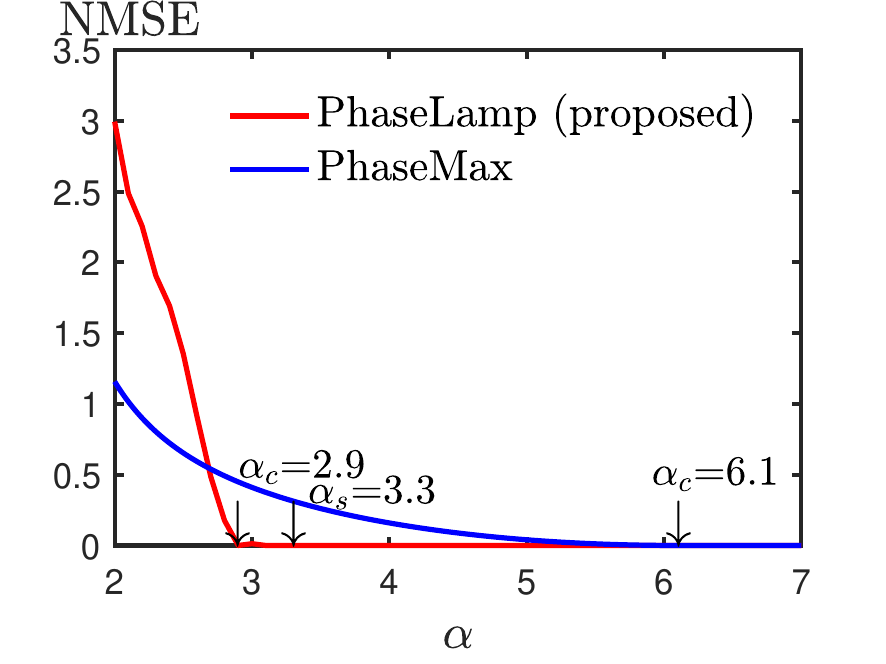}}

    \caption{Sufficient condition for successful recovery of the PhaseLamp method. The NMSE as a function of the oversampling ratio $\alpha$, for (a) $\rho_{\text{init}}=0.1$; (b) $\rho_{\text{init}}=0.3$. The maximum number of iterations is set to $I_{\text{max}}=20$ and $\epsilon=10^{-4}$. The results are averaged over $10$ independent Monte Carlo trials. In both cases, the PhaseLamp method outperforms the PhaseMax method. Moreover, the sufficient condition for perfect recovery \eref{slamp} denoted by $\alpha_s$ provides a good estimate of the actual recovery performance of the PhaseLamp method.}
    \label{fig:fig2}
\end{figure}

\section{Conclusion}
\label{sec:conclusion}

We presented in this paper an asymptotically exact characterization of the performance of the PhaseMax method for phase retrieval. Specifically, our analysis reveals a sharp phase transition behavior in the performance of the method as one varies the oversampling ratio and the input cosine similarity. Our analysis is based on the CGMT, and the results match previous predictions derived from the non-rigorous replica method. Moreover, we also presented a new nonconvex formulation of the phase retrieval problem and PhaseLamp, an iterative algorithm based on  linearization and maximization over a polytope. We provided a sufficient condition for PhaseLamp to perfectly retrieve the target vector. Simulation results confirm the validity of our theoretical predictions. They also show that the proposed iterative algorithm significantly improves the recovery performance of the original PhaseMax method.


\appendix
In this appendix we provide proof sketches for Theorems \ref{thm:the1} and \ref{thm:pt}.
\subsection{Notation}

To simplify the exposition we assume onwards that $\vxi=\ve_1$, the first vector of the canonical basis of $\mathbb{R}^n$, and that $\norm{\vx_\text{init}}_2 = 1$. This assumption can be made without loss of generality due to rotational invariance of the Gaussian distribution and since the optimization problems \eref{lp_form} and \eref{qb_form} are scale invariant.

Further, we stack the sensing vectors $\lbrace \va_i^T, 1 \leq i \leq m \rbrace$ to form the sensing matrix $\mA \in\mathbb{R}^{m\times n}$. Let $\vq$ denote the first column of $\mA$ and $\mG\in\mathbb{R}^{m\times(n-1)}$ the remaining part, \emph{i.e.}, $$\mA=[\vq~~\mG].$$ Similarly, define $\eta_1$ and $\widetilde{\veta}$ such that $$\vx_{\text{init}}=[\eta_1~~\widetilde{\veta}^T]^T.$$ Finally, for a vector $\vc$, we let  $\abs{\vc}$ and $\text{sign}(\vc)$ to denote its component-wise absolute value and sign, respectively. Also, we let $\min(\vc)$ return the minimum value in the vector, and, $\vz=\vc\wedge\mathbf{0}$ be a vector such that $\vz_i=\min(\vc_i,0)$. 
\subsection{Convex Gaussian Min-Max Theorem (CGMT)}

The proof follows the CGMT framework\cite{chris:151,chris:152}. For ease of reference we summarize here the essential ideas of the framework; please see \cite[Section~6]{chris:151} for the formal statement of the theorem and further details.
The CGMT associates with a primary optimization (PO) problem a simplified auxiliary optimization (AO) problem from which we can tightly infer properties of the original (PO), including the optimal cost and the optimal solution. The two problems are of the following form:
\begin{equation}\label{eq:PO}
\Phi(\mC)=\min\limits_{\vw \in \mathcal{S}_{\vw}} \max\limits_{\vu\in\mathcal{S}_{\vu}} \vu^T \mC \vw + \psi(\vw,\vu),
\end{equation}
and
\begin{equation}\label{eq:AO}
\hspace{-1mm}\phi(\vg,\vh)=\min\limits_{\vw \in \mathcal{S}_{\vw}} \max\limits_{\vu\in\mathcal{S}_{\vu}} \norm{\vw}_2 \vg^T \vu + \norm{\vu}_2 \vh^T \vw + \psi(\vw,\vu),
\end{equation}
respectively. Above, $\mC\in\mathbb{R}^{m\times n}$, $\vg \in \mathbb{R}^m$, $\vh\in\mathbb{R}^n$, the sets $\mathcal{S}_{\vw}\subset\R^n$ and $\mathcal{S}_{\vu}\subset\R^m$ are compact, and, $\psi: \mathbb{R}^n\times \mathbb{R}^m \to \mathbb{R}$. We think of the two problems in \eqref{eq:PO} and \eqref{eq:AO} as random optimization problems, in which $\mC$, $\vg$ and $\vh$ all have i.i.d standard normal entries. According to the CGMT \cite[Theorem 6.1]{chris:151}, if the sets $\mathcal{S}_{\vw}$ and $\mathcal{S}_{\vu}$ are convex and $\psi$ is continuous \emph{convex-concave} on $\mathcal{S}_{\vw}\times \mathcal{S}_{\vu}$, then, for any $\mu \in \mathbb{R}$ and $t>0$, it holds
\begin{equation}\label{eq:cgmt}
\mathbb{P}\left( \abs{\Phi(\mC)-\mu} > t\right) \leq 2 \mathbb{P}\left(  \abs{\phi(\vg,\vh)-\mu} > t \right).
\end{equation}
In words, concentration of the optimal cost of the AO problem around $\mu$ implies concentration of the optimal cost of the corresponding PO problem around the same value $\mu$. Moreover, starting from \eqref{eq:cgmt} and under strict convexity conditions, the CGMT shows that concentration of the optimal solution of the AO problem implies concentration of the optimal solution of the PO to the same value. For example, if minimizers of \eqref{eq:AO} satisfy $\norm{\vw^\ast(\vg,\vh)}_2 \to \zeta^\ast$ for some $\zeta^\ast>0$, then, the same holds true for the minimizers of \eqref{eq:PO}: $\norm{\vw^\ast(\mC)}_2 \to \zeta^\ast$ \cite[Theorem 6.1(iii)]{chris:151}. Thus, one can analyze the AO to infer corresponding properties of the PO, the premise being of course that the former is simpler to handle than the latter.

\subsection{CGMT for the PhaseMax Method}
We apply the CGMT to characterize the asymptotic NMSE of the PhaseMax optimization in \eqref{eq:lp_form} as in \eqref{eq:anmse} and \eref{phb}. In this section, we write the PhaseMax optimization in the form of a PO as in \eqref{eq:PO}, which in turn leads to a corresponding AO optimization problem. For these problems, we can show that the conditions of the CGMT on convexity and compactness are satisfied.

First, we appropriately write the linear program in \eqref{eq:lp_form} as a minmax program. Start with its dual: 
\begin{align}\label{eq:po_form}
\hspace{-1mm}\min_{\substack{ \vlambda\geq 0 \\ \vmu\geq 0 }} \vmu^T \abs{\mA \vxi} + \vlambda^T  \abs{\mA \vxi} ~~\text{s.t.}~~ \mA^T\vmu- \mA^T\vlambda={\vx}_\text{init}.
\end{align}
Since $m>n$ (recall: $\alpha>2$) both the primal and the dual are bounded feasible 
with probability one, and strong duality holds \cite{BoydV:03}.
Therefore, \eref{po_form} is equivalent to the following
\begin{align}\label{eq:pof_form}
\hspace{-3mm}\min_{\substack{ \vlambda\in\mathcal{S}_{\vlambda} \\ \vmu\in\mathcal{S}_{\vmu} }} \underset{{\vx}}{\max}~{\vx}_\text{init}^{T}\,{\vx}+\vmu^T \left(\, \abs{\mA \vxi}-\mA \vx \right)+\vlambda^T \left(\, \abs{\mA \vxi}+\mA \vx \right),
\end{align}
for $\mathcal{S}_{\vlambda}=\lbrace \vlambda\in\mathbb{R}^m~\text{s.t}~\norm{\vlambda}_{\infty} \leq \Gamma,~\vlambda \geq 0 \rbrace$, $\mathcal{S}_{\vmu}=\lbrace \vmu\in\mathbb{R}^m~\text{s.t}~\norm{\vmu}_{\infty} \leq \Gamma,~\vmu \geq 0 \rbrace$, and $\Gamma$ a sufficiently large positive constant. Note that the feasibility sets $\mathcal{S}_{\vlambda}$ and $\mathcal{S}_{\vmu}$ are convex and compact in $\mathbb{R}^m$, the feasibility set of the variable $\vx$ is convex and the cost function in \eref{pof_form} is linear with respect to each optimization variable. Thus, the order of min-max can be flipped to obtain the following equivalent,
\begin{align}\label{eq:poff_form}
\hspace{-2mm} \underset{{\vx}}{\max} \min_{\substack{ \vlambda\in\mathcal{S}_{\vlambda} \\ \vmu\in\mathcal{S}_{\vmu} }} {\vx}_\text{init}^{T}\,{\vx}+\vmu^T \left(\, \abs{\mA \vxi}-\mA \vx \right)+\vlambda^T \left(\, \abs{\mA \vxi}+\mA \vx \right).
\end{align}
From strict duality, this is also equivalent to \eqref{eq:lp_form}. Furthermore, as already discussed, the optimality set of \eqref{eq:lp_form} is bounded. Naturally then there is a sufficiently large constant $B>0$ and sets $\mathcal{S}_{x_1}=\lbrace x_1 \in\mathbb{R}~\text{s.t}~|x_1| \leq B \rbrace$ and $\mathcal{S}_{\widetilde{\vx}}=\lbrace \widetilde{\vx}\in\mathbb{R}^{n-1}~\text{s.t}~\norm{\widetilde{\vx}}_{\infty} \leq B \rbrace$ such that \eqref{eq:poff_form} can be formulated as
%
\begin{align}\label{eq:poff_form_2}
\hspace{-2mm}\underset{\substack{ {x}_1\in\mathcal{S}_{x_1}\\ \widetilde{\vx}\in\mathcal{S}_{\widetilde{\vx}} }}{\max} \min_{\substack{ \vlambda\in\mathcal{S}_{\vlambda} \\ \vmu\in\mathcal{S}_{\vmu} }}&  \ \eta_1 x_1 + \widetilde{\veta}^T \widetilde{\vx} +(\vlambda-\vmu)^T \vq x_1 + (\vlambda-\vmu)^T \mG \widetilde{\vx}\nonumber\\
&\qquad + (\vlambda+\vmu)^T \abs{\vq}
\end{align}
At this point, observe that  \eqref{eq:poff_form_2} is in the desired form of a PO as in \eqref{eq:PO} with $\mG\in\R^{m\times (n-1)}$ having i.i.d standard normal entries and the function $\psi$, defined as 
\begin{equation}
\psi(\vx,(\vlambda,\vmu))=\eta_1 x_1 + \widetilde{\veta}^T \widetilde{\vx} +(\vlambda-\vmu)^T \vq x_1 + (\vlambda+\vmu)^T \abs{\vq}.\notag
\end{equation}
Further note that the constraint sets are convex compact and $\psi$ is concave-convex on $\mathcal{S}_{\vx}\times (\mathcal{S}_{\vlambda}\times \mathcal{S}_{\vmu})$, where $\vx=[x_1~\widetilde{\vx}^T]^T$ and $\mathcal{S}_{\vx}=\mathcal{S}_{x_1}\times\mathcal{S}_{\widetilde{\vx}}$.

We are now ready to formulate the corresponding AO problem:
\begin{align}\label{eq:ao_form}
\hspace{-2mm}\underset{\substack{ {x}_1\in\mathcal{S}_{x_1}\\ \widetilde{\vx}\in\mathcal{S}_{\widetilde{\vx}} }}{\max} \min_{\substack{ \vlambda\in\mathcal{S}_{\vlambda} \\ \vmu\in\mathcal{S}_{\vmu} }}&\   \norm{\widetilde{\vx}}_2 \vg^T (\vlambda-\vmu) + \norm{\vlambda-\vmu}_2 \vh^T \widetilde{\vx} + \eta_1 x_1  \nonumber\\
& + \widetilde{\veta}^T \widetilde{\vx}  +(\vlambda-\vmu)^T \vq x_1 + (\vlambda+\vmu)^T \abs{\vq}.
\end{align}
Following the CGMT framework we proceed onwards with analyzing \eqref{eq:ao_form}.

\subsection{Analysis of the Auxiliary Optimization Problem}
\vsp
\subsubsection{Simplifying the AO}
Consider the following change of variables: $\vv=\vlambda-\vmu$ and $\vb=\vlambda+\vmu$. To respect the nonnegativity of  $\vlambda$ and $\vmu$, it must be that $\vb\geq|\vv|$. In fact, it can be checked that the optimal solution for $\vb$ is $\vb=|\vv|$. Thus, the optimization problem \eref{ao_form} can be reduced to the following:
\begin{align}
\hspace{-2mm}\underset{\substack{ {x}_1\in\mathcal{S}_{x_1}\\ \widetilde{\vx}\in\mathcal{S}_{\widetilde{\vx}} }}{\max} \min_{ \vv\in\mathcal{S}_{\vv} }&  \left( \norm{\widetilde{\vx}}_2 \vg + x_1\vq \right)^T \vv + \norm{\vv}_2 \vh^T \widetilde{\vx} + \eta_1 x_1  \nonumber\\ \notag
& + \widetilde{\veta}^T \widetilde{\vx}  + \abs{\vv}^T \abs{\vq}.
\end{align}
Above, the set $\mathcal{S}_{\vv}$ is defined as $\mathcal{S}_{\vv}=\lbrace \vv \in\mathbb{R}^{m}~\text{s.t}~\norm{\vv}_{\infty} \leq \Delta \rbrace$, where $\Delta$ is a sufficiently large positive constant. 

Next, observe that if we fix $\abs{\vv}$, then the optimal $\vv$ satisfies $\text{sign}(\vv)=-\text{sign}\left(\norm{\widetilde{\vx}}_2 \vg + \vq x_1\right)$ which simplifies the optimization to the following
\begin{align}
\hspace{-2mm}\underset{\substack{ {x}_1\in\mathcal{S}_{x_1}\\ \widetilde{\vx}\in\mathcal{S}_{\widetilde{\vx}} }}{\max} \min_{ \Delta\geq\vv\geq 0 }& [|\vq| -|\norm{\widetilde{\vx}}_2 \vg + \vq x_1|]^T\vv  +  \norm{\vv}_2 \vh^T \widetilde{\vx} \nonumber\\
&+ \eta_1 x_1 + \widetilde{\veta}^T \widetilde{\vx}. \notag
\end{align}
%
Next, in the optimization above one can fix the norm of $\vv$  and optimize over its direction. Omitting some details, the optimization becomes
\begin{align}\label{eq:ao2_form}
&\underset{ \substack{ {x}_1\in\mathcal{S}_{x_1} \\ \widetilde{\vx}\in\mathcal{S}_{\widetilde{\vx}} } }{\max}~\eta_1 x_1 + \widetilde{\veta}^T \widetilde{\vx}  \\
&~~~\text{s.t.}~~ \vh^T \widetilde{\vx} + h\left( \abs{\vq} - \abs{ \norm{\widetilde{\vx}}_2 \vg + x_1 \vq}  \right) \geq 0, \nonumber 
\end{align}
where we defined the function $h:\R^n\rightarrow\R$ as 
\begin{equation}
h(\vc)=\begin{cases}
-\norm{\vc\wedge\mathbf{0}}_2 & ,\text{if}~\min(\vc) \leq 0,\notag \\
\min(\vc) & ,\text{otherwise}. \notag
\end{cases}
\end{equation}

The final step in simplifying the AO problem is as follows. For fixed value of $x_1$ (say $x_1=s>0$), and for fixed norm of $\widetilde{\vx}$ (say, $\norm{\widetilde{\vx} }_2=r$), we optimize over the direction of $\widetilde{\vx}$.
It can be shown that this optimization further reduces \eqref{eq:ao2_form} to the following two-dimensional optimization problem:
\begin{align}\label{eq:ao3_form}
&\underset{ \substack{ s\in\mathcal{S}_{x_1} \\ B\geq r\geq 0 } }{\max}~\eta_1 s + \norm{\widetilde{\veta}}_2 \sqrt{r^2-c(s,r)^2} \\
&~~~\text{s.t.}~~ c(s,r) \leq r, \nonumber 
\end{align}
where the function $c:\R\times\R\rightarrow\R$ is defined as 
\begin{equation}\notag
c(s,r)=-\frac{h\left( \abs{\vq} - \abs{ r \vg + s \vq}  \right)}{\norm{\vh}_2}.
\end{equation}

\vsp
\subsubsection{Convergence Analysis}
Now that we have simplified the AO to a maximization problem over only two scalar variables as in \eqref{eq:ao3_form}, we are ready to study its asymptotic behavior in the regime $m,n\rightarrow\infty, m/n\rightarrow\alpha$.
 Specifically, it can be shown that the optimization problem \eref{ao3_form} converges to the following \emph{deterministic} optimization problem
\begin{align}\label{eq:aod_form}
&\underset{ \substack{ |s|<B \\ B\geq r\geq 0 } }{\max}~\eta_1 s + \norm{\widetilde{\veta}}_2 \sqrt{r^2-\alpha~c_d(r,s)  } \\
&~~~\text{s.t.}~~ c_d(s,r) \leq r^2/\alpha, \nonumber 
\end{align}
where $c_d(s,r)=\mathbb{E}_{q, g} \big[ \min\big\{\abs{q} - \abs{ r g + s q },0\big\}^2 \big]$, which takes the following closed-form:
\begin{align}
&c_d(s,r)=\frac{1}{\pi} \Bigg[  ( (1-s)^2+r^2 )\left(\frac{\pi}{2}-\text{atan}\left( \frac{1-s}{r} \right) \right)\nonumber\\
&+( (1+s)^2+r^2 )\left(\frac{\pi}{2}-\text{atan}\left( \frac{1+s}{r} \right) \right)-2r \Bigg].
\end{align}
The full technical details of obtaining the convergence result in \eref{aod_form} are deferred to the full version of the paper. In short, pointwise convergence of the objective function of \eqref{eq:ao3_form} to \eqref{eq:aod_form} for fixed $s$ and $r$ follows easily from the weak law of large numbers. The corresponding convergence of the optimal costs requires proof of uniform convergence, which follows by pointwise convergence and concavity of the objective function \cite[Lemma~7.75]{chakraborty2008statistical}. 
We call the deterministic two-dimensional optimization problem in \eqref{eq:aod_form} as the scalar performance optimization (SPO); according to the CGMT solving the SPO allows us to conclude on the asymptotic performance of the PhaseMax problem (cc. the PO). 

\vsp

\subsubsection{Solving the scalar performance optimization}
Recall that the SPO in \eqref{eq:aod_form} is the converging limit of the AO in \eqref{eq:ao_form}. Specifically, the optimization variables $s$ and $r$ in \eqref{eq:aod_form}  correspond exactly to  $x_1$ and $\|\widetilde\vx\|_2$ in \eqref{eq:ao_form}. From this and uniform convergence discussed previously, the optimal values of $s$ and $r$ are the converging limits of $x_1$ and of $\|\widetilde\vx\|_2$, respectively. In what follows, we solve the SPO problem for the optimal $s$ and $r$.
First, using the assumption of the theorem that $\alpha > 2$, it can be shown that  the feasible set of \eqref{eq:aod_form} is nonempty iff $\abs{s} \leq 1$.
Second, for fixed $|s|\leq 1$, the problem
\begin{align}\label{eq:aod2_form}
&\underset{  r\geq 0}{\max}~ r^2-\alpha\,c_d(s,r) 
\end{align} 
is concave and admits a unique solution 
\begin{equation}\label{eq:vstar}
r^\ast(s)=\sqrt{\frac{1}{\text{tan}\left( \frac{\pi}{\alpha} \right)^2}+(1-s^2)}-\frac{1}{\text{tan}\left( \frac{\pi}{\alpha} \right)}.
\end{equation}
At this point, note that we can always find a large enough constant $\widetilde B>0$ such that $r^\ast(s)<\widetilde B$ for all $|s|<1$. Therefore, choosing $B$ in \eqref{eq:aod_form} such that $B=\widetilde B$ guarantees that the optimal value of $r$ in \eqref{eq:aod_form} is given by \eqref{eq:vstar}. Substituting this value back in \eqref{eq:aod_form}, we can now optimize over $s$ by solving the following:
\begin{align}\label{eq:aod4_form}
&{\max_{|s|\leq 1}}~\eta_1 s + \norm{\widetilde{\veta}}_2 \sqrt{(r^\ast(s))^2-\alpha~c_d(r^\ast(s),s)  }.
\end{align}
A few algebra manipulations show that \eqref{eq:aod4_form} is equivalent to \eref{fconprob} in the statement of the theorem. To show the equivalence, further note that $\eta_1$ and $\widetilde{\veta}$ in \eqref{eq:aod4_form} are related to the input cosine similarity $\rho_{\text{init}}$, defined in \eref{cosin}, as follows (recall: $\boldsymbol{\xi}=\ve_1$.),
\begin{equation}
\frac{\eta_1}{\norm{\widetilde{\veta}}_2} = \frac{\rho_{\text{init}}}{\sqrt{1-\rho_{\text{init}}^2}}.
\end{equation}
Finally, note that the optimization in \eqref{eq:aod4_form} [eqv., in \eqref{eq:fconprob}] inherits the concavity of \eqref{eq:aod_form}, \emph{i.e.}, it is a concave program.

\vsp
\subsubsection{Phase transition calculations}
In this section, we compute the phase transition boundary of the PhaseMax method. Our goal is to find necessary and sufficient conditions under which the solution $\hat\vx$ of the PhaseMax is, with high probability, equal to $\boldsymbol{\xi}=\ve_1$. Mapping this to the SPO in \eqref{eq:aod4_form} [eqv., see \eqref{eq:fconprob}], we seek conditions under which $s^\ast=1$ and $r^\ast=0$. From concavity, this happens if and only if the 
derivative of the cost function of the optimization problem \eref{fconprob} at $s=1$ is nonnegative.  
%
 Hence, the necessary and sufficient condition for perfect recovery of the PhaseMax method is given by 
\begin{equation}
\frac{\rho_{\text{init}}}{\sqrt{1-\rho_{\text{init}}^2}} > \sqrt{\frac{\alpha}{\pi}\text{tan}\left(  \frac{\pi}{\alpha}\right)-1},
\end{equation}
for $\alpha > 2$. 
Equivalently, the oversampling ratio $\alpha$ and the input cosine similarity given in \eref{cosin} must satisfy
\begin{equation}\label{eq:ptb}
\frac{\pi}{\alpha~\text{tan}\left(  \pi / \alpha \right)} > 1-\rho_{\text{init}}^2.
\end{equation}
This then gives us the statement of Theorem \ref{thm:pt}.


\bibliographystyle{IEEEtran}
\bibliography{reference,refs}

\begin{thebibliography}{10}
\providecommand{\url}[1]{#1}
\csname url@samestyle\endcsname
\providecommand{\newblock}{\relax}
\providecommand{\bibinfo}[2]{#2}
\providecommand{\BIBentrySTDinterwordspacing}{\spaceskip=0pt\relax}
\providecommand{\BIBentryALTinterwordstretchfactor}{4}
\providecommand{\BIBentryALTinterwordspacing}{\spaceskip=\fontdimen2\font plus
\BIBentryALTinterwordstretchfactor\fontdimen3\font minus
  \fontdimen4\font\relax}
\providecommand{\BIBforeignlanguage}[2]{{%
\expandafter\ifx\csname l@#1\endcsname\relax
\typeout{** WARNING: IEEEtran.bst: No hyphenation pattern has been}%
\typeout{** loaded for the language `#1'. Using the pattern for}%
\typeout{** the default language instead.}%
\else
\language=\csname l@#1\endcsname
\fi
#2}}
\providecommand{\BIBdecl}{\relax}
\BIBdecl

\bibitem{Gerchberg:1972jk}
R.~W. Gerchberg, ``A practical algorithm for the determination of phase from
  image and diffraction plane pictures,'' \emph{Optik}, vol.~35, p. 237, 1972.

\bibitem{Fienup:82}
J.~R. Fienup, ``Phase retrieval algorithms: a comparison,'' \emph{Applied
  Optics}, vol.~21, no.~15, pp. 2758--2769, 1982.

\bibitem{Candes:2013xy}
E.~J. Candes, T.~Strohmer, and V.~Voroninski, ``Phaselift: {Exact} and stable
  signal recovery from magnitude measurements via convex programming,''
  \emph{Communications on Pure and Applied Mathematics}, vol.~66, no.~8, pp.
  1241--1274, 2013.

\bibitem{Jaganathan:2013zl}
K.~Jaganathan, S.~Oymak, and B.~Hassibi, ``Sparse phase retrieval: {Convex}
  algorithms and limitations,'' in \emph{Information {Theory} {Proceedings}
  ({ISIT}), 2013 {IEEE} {International} {Symposium} on}.\hskip 1em plus 0.5em
  minus 0.4em\relax IEEE, 2013, pp. 1022--1026.

\bibitem{Waldspurger:2015rz}
I.~Waldspurger, A.~d'Aspremont, and S.~Mallat, ``Phase recovery, maxcut and
  complex semidefinite programming,'' \emph{Mathematical Programming}, vol.
  149, no. 1-2, pp. 47--81, 2015.

\bibitem{Netrapalli:2013qv}
P.~Netrapalli, P.~Jain, and S.~Sanghavi, ``Phase retrieval using alternating
  minimization,'' in \emph{Advances in {Neural} {Information} {Processing}
  {Systems}}, 2013, pp. 2796--2804.

\bibitem{Candes:2015fv}
E.~J. Candes, X.~Li, and M.~Soltanolkotabi, ``Phase retrieval via {Wirtinger}
  flow: {Theory} and algorithms,'' \emph{Information Theory, IEEE Transactions
  on}, vol.~61, no.~4, pp. 1985--2007, 2015.

\bibitem{WangGY:2016}
G.~Wang, G.~B. Giannakis, and Y.~C. Eldar, ``Solving {Systems} of {Random}
  {Quadratic} {Equations} via {Truncated} {Amplitude} {Flow},''
  \emph{arXiv:1605.08285}, May 2016.

\bibitem{balan2009painless}
R.~Balan, B.~G. Bodmann, P.~G. Casazza, and D.~Edidin, ``Painless
  reconstruction from magnitudes of frame coefficients,'' \emph{Journal of
  Fourier Analysis and Applications}, vol.~15, no.~4, pp. 488--501, 2009.

\bibitem{Chen:2015eu}
Y.~Chen and E.~J. Candes, ``Solving {Random} {Quadratic} {Systems} of
  {Equations} {Is} {Nearly} as {Easy} as {Solving} {Linear} {Systems},''
  \emph{arXiv:1505.05114}, 2015.

\bibitem{LuL:17}
\BIBentryALTinterwordspacing
Y.~M. Lu and G.~Li, ``Phase transitions of spectral initialization for
  high-dimensional nonconvex estimation,'' \emph{arXiv:1702.06435 [cs.IT]},
  2017. [Online]. Available: \url{https://arxiv.org/abs/1702.06435}
\BIBentrySTDinterwordspacing

\bibitem{phmax2}
\BIBentryALTinterwordspacing
S.~Bahmani and J.~Romberg, ``{Phase Retrieval Meets Statistical Learning
  Theory: {A} Flexible Convex Relaxation},'' \emph{CoRR}, vol. abs/1610.04210,
  2016. [Online]. Available: \url{http://arxiv.org/abs/1610.04210}
\BIBentrySTDinterwordspacing

\bibitem{phmax}
\BIBentryALTinterwordspacing
T.~Goldstein and C.~Studer, ``{{PhaseMax}: Convex Phase Retrieval via Basis
  Pursuit},'' \emph{CoRR}, vol. abs/1610.07531, 2016. [Online]. Available:
  \url{http://arxiv.org/abs/1610.07531}
\BIBentrySTDinterwordspacing

\bibitem{Hand:2016cs}
\BIBentryALTinterwordspacing
P.~Hand and V.~Voroninski, ``An {Elementary} {Proof} of {Convex} {Phase}
  {Retrieval} in the {Natural} {Parameter} {Space} via the {Linear} {Program}
  {PhaseMax},'' \emph{arXiv:1611.03935 [cs, math]}, Nov. 2016, arXiv:
  1611.03935. [Online]. Available: \url{http://arxiv.org/abs/1611.03935}
\BIBentrySTDinterwordspacing

\bibitem{Ouss:17}
\BIBentryALTinterwordspacing
O.~{Dhifallah} and Y.~M. {Lu}, ``{Fundamental Limits of PhaseMax for Phase
  Retrieval: A Replica Analysis},'' \emph{Proc. International Workshop on
  Computational Advances in Multi-Sensor Adaptive Processing (CAMSAP)}, 2017.
  [Online]. Available: \url{https://arxiv.org/abs/1708.03355}
\BIBentrySTDinterwordspacing

\bibitem{chris:151}
\BIBentryALTinterwordspacing
C.~Thrampoulidis, E.~Abbasi, and B.~Hassibi, ``Precise error analysis of
  regularized m-estimators in high-dimensions,'' \emph{CoRR}, vol.
  abs/1601.06233, 2016. [Online]. Available:
  \url{http://arxiv.org/abs/1601.06233}
\BIBentrySTDinterwordspacing

\bibitem{chris:152}
C.~Thrampoulidis, S.~Oymak, and B.~Hassibi, ``Regularized linear regression: A
  precise analysis of the estimation error,'' in \emph{Proceedings of The 28th
  Conference on Learning Theory}, vol.~40.\hskip 1em plus 0.5em minus
  0.4em\relax Paris, France: PMLR, 03--06 Jul 2015, pp. 1683--1709.

\bibitem{Gordon:85}
Y.~Gordon, ``Some inequalities for {Gaussian} processes and applications,''
  \emph{Israel Journal of Mathematics}, vol.~50, no.~4, pp. 265--289, Dec 1985.

\bibitem{Sto:13}
\BIBentryALTinterwordspacing
M.~Stojnic, ``A framework to characterize performance of {LASSO} algorithms,''
  \emph{CoRR}, vol. abs/1303.7291, 2013. [Online]. Available:
  \url{http://arxiv.org/abs/1303.7291}
\BIBentrySTDinterwordspacing

\bibitem{thrampoulidis2015asymptotically}
C.~Thrampoulidis, A.~Panahi, and B.~Hassibi, ``Asymptotically exact error
  analysis for the generalized equation-lasso,'' in \emph{Information Theory
  (ISIT), 2015 IEEE International Symposium on}.\hskip 1em plus 0.5em minus
  0.4em\relax IEEE, 2015, pp. 2021--2025.

\bibitem{thrampoulidis2015lasso}
C.~Thrampoulidis, E.~Abbasi, and B.~Hassibi, ``Lasso with non-linear
  measurements is equivalent to one with linear measurements,'' in
  \emph{Advances in Neural Information Processing Systems}, 2015, pp.
  3420--3428.

\bibitem{hand2016elementary}
P.~Hand and V.~Voroninski, ``An elementary proof of convex phase retrieval in
  the natural parameter space via the linear program phasemax,''
  \emph{arXiv:1611.03935}, 2016.

\bibitem{fasta}
\BIBentryALTinterwordspacing
T.~Goldstein, C.~Studer, and R.~Baraniuk, ``A field guide to forward-backward
  splitting with a {FASTA} implementation,'' \emph{arXiv eprint}, vol.
  abs/1411.3406, 2014. [Online]. Available:
  \url{http://arxiv.org/abs/1411.3406}
\BIBentrySTDinterwordspacing

\bibitem{Lange:16}
K.~Lange, \emph{{MM} Optimization Algorithms}.\hskip 1em plus 0.5em minus
  0.4em\relax {SIAM}, 2016.

\bibitem{BoydV:03}
S.~Boyd and L.~Vandenberghe, \emph{Convex Optimization}.\hskip 1em plus 0.5em
  minus 0.4em\relax Cambridge University Press, 2003.

\bibitem{chakraborty2008statistical}
K.-J. Miescke and F.~Liese, \emph{Statistical Decision Theory: Estimation,
  Testing, and Selection}.\hskip 1em plus 0.5em minus 0.4em\relax Springer New
  York, 2008.

\end{thebibliography}

\end{document}